\pdfoutput=1
\def\ts     {\thinspace}
\def\kms    {\ifmmode{{\rm \ts km\ts s}^{-1}}\else{\ts km\ts s$^{-1}$}\fi}
\def\msol   {\ifmmode{{\rm M}_{\odot} }\else{M$_{\odot}$}\fi}
\def\lsol   {\ifmmode{L_{\odot}}\else{$L_{\odot}$}\fi}
\def\lfir   {\ifmmode{L_{\rm FIR}}\else{$L_{\rm FIR}$}\fi}
\def\zsol   {\ifmmode{{\rm Z}_{\odot}}\else{Z$_{\odot}$}\fi}
\def\etal   {{\rm et\ts al.\ }}
\def\ci     {\ifmmode{{\rm C}{\rm \small I}}\else{C\ts {\scriptsize I}}\fi}
\def\hi     {\ifmmode{{\rm H}{\rm \small I}}\else{H\ts {\scriptsize I}}\fi}
\def\hh     {\ifmmode{{\rm H}_2}\else{H$_2$}\fi}
\def\cone {\ifmmode{{\rm C}{\rm \small I}(^3\!P_1\!\to^3\!P_0)}
     \else{C\ts {\scriptsize I}{\small$(^3\!P_1\!\to^3\!\!\!P_0)$}}\fi}
\def\ctwo {\ifmmode{{\rm C}{\rm \small I}(^3\!P_2\!\to^3\!P_1)}
     \else{C\ts {\scriptsize I}{\small$(^3\!P_2\!\to^3\!\!\!P_1)$}}\fi}
\def\cij {\ifmmode{{\rm C}{\rm \small I}\,(^3P_i\to^3P_j)}
\else{C\ts {\scriptsize I}\,{\small$(^3P_i\to^3P_j)$}}\fi}
\def\cii    {\ifmmode{{\rm C}{\rm \small II}}\else{C\ts {\scriptsize II}}\fi}
\def\tex {\ifmmode{{T}_{\rm ex}}\else{$T_{\rm ex}$}\fi}
\def\tmb {\ifmmode{{T}_{\rm mb}}\else{$T_{\rm mb}$}\fi}
\def\tkin {\ifmmode{{T}_{\rm kin}}\else{$T_{\rm kin}$}\fi}
\def\microns {\ifmmode{\mu{\rm m}}\else{$\mu$m}\fi}
\def\nhh   {\ifmmode{n({\rm H}_2)}\else{$n$(H$_2$)}\fi}
\def\gradv {\ifmmode{(dv/dr)}\else{$(dv/dr)$}\fi}
\def\CO10{{\hbox {CO(1--0)}}}

\def\,{\thinspace}
\def\Msun{M$_\odot$}

\def\Lsun{L$_\odot$}

\def \Kkmspc{K\,\kms\,pc$^2$}

\documentclass[twocolumn,bibyear]{aa}    
\usepackage{graphicx}
\begin{document}
\titlerunning{C$^+$ 158\,$\mu$m line in the HDF\,850.1 merger galaxies}
\title{High-resolution C$^+$ imaging of HDF\,850.1 reveals a merging galaxy at z=5.185}
   \author{
          R. Neri
          \inst{1}
          \and
          D. Downes
          \inst{1}
          \and
          P. Cox
          \inst{1,2}
          \and
          F. Walter
          \inst{3}
          }
   \institute{Institut de Radio Astronomie Millim\'etrique,
              Domaine Universitaire, 38406 St-Martin-d'H\`eres, France
         \and
   ALMA SCO, Alonso de Cordova 3107, Vitacura, Santiago, Chile
         \and
   Max-Planck-Institut f\"ur Astronomie,
           K\"onigstuhl 17, D-69117 Heidelberg, Germany
             }

   \date{Received 22 August 2013 / Accepted 12 December 2013}
   \abstract{
New high-resolution maps with the IRAM Interferometer of the
redshifted C$^+$ 158\,$\mu$m line and the 0.98\,mm dust continuum of
HDF\,850.1 at $z$ = 5.185 show the source to have a blueshifted
northern component and a redshifted southern component, with a
projected separation of 0.3$''$, or 2\,kpc.  We interpret these
components as primordial galaxies that are merging to form a larger
galaxy.   We think it is the resulting merger-driven starburst
that makes HDF\,850.1 an ultraluminous infrared galaxy, with $L_{IR} \sim
10^{13}$\,\Lsun .
The observed line and continuum brightness temperatures and the
constant line-to-continuum ratio across the source imply (1) high
C$^+$ line optical depth, (2) a C$^+$ excitation temperature of
the same order as the dust temperature, and (3) dust continuum
emission that is nearly optically thick at 158\,$\mu$m.
These conclusions for HDF\,850.1 probably also apply to other
high-redshift submillimeter galaxies and quasar hosts in which
the C$^+$ 158\,$\mu$m line has been detected, as indicated by their
roughly constant C$^+$-to-158\,$\mu$m continuum ratios, in sharp
constrast to the large dispersion in their C$^+$-to-FIR luminosity ratios.
In brightness temperature units, the C$^+$ line luminosity is about
the same as the predicted CO(1--0) luminosity, implying that the
C$^+$ line can also be used to estimate the molecular gas mass,
with the same assumptions as for CO.

}
   \maketitle
\section{Introduction}
HDF\,850.1, the brightest submillimeter
galaxy (SMG) in the Hubble Deep Field, has been shown by Walter et al.\ (2012)
to have redshift of $z$ = 5.2, and to lie in a group of
at least 12 galaxies in a redshift bin from $z$ =5.18 to 5.21, in the
GOODS-N field containing the Hubble Deep Field.
We now continue the observations of
Walter et al.\ (2012; see that paper for references
to earlier multi-wavelength data on HDF\,850.1), and
report follow-up, higher-resolution maps in the
redshifted C$^+$ 158$\,\mu$m line and the 0.98\,mm dust continuum.
Walter et al. presented a rotating disk interpretation of the C$^+$ data.
We now present an alternative interpretation, regarding the new observations
as evidence for a galaxy merger.
We also use the new data to update our previous model (Walter et al.\ 2012)
for the gravitational lensing
by the $z$ = 1.2 elliptical galaxy 3-586.0.

The higher resolution enables us, for the first time for HDF\,850.1,
to measure the dust brightness temperature, and hence to estimate the
dust optical depth at rest-frame 158$\,\mu$m.  These new
observations also allow us to better resolve the C$^+$ line, and hence
to map for the first time the C$^+$-to-dust flux density ratio across
the source, which turns out to be nearly constant.  We
compare this result with the C$^+$-to-dust ratio in other
high-redshift galaxies, and discuss the interpretation,
on kpc scales, of C$^+$ at high-$z$, and the
possible use of the C$^+$ line to estimate the gas mass.

In the light of these new observations, we discuss why there is as yet no
rest-frame optical or UV counterpart to HDF\,850.1, and we summarize
the evidence for our merger interpretation.

\begin{figure*} \centering
\includegraphics[angle=-0,width=8.5cm]{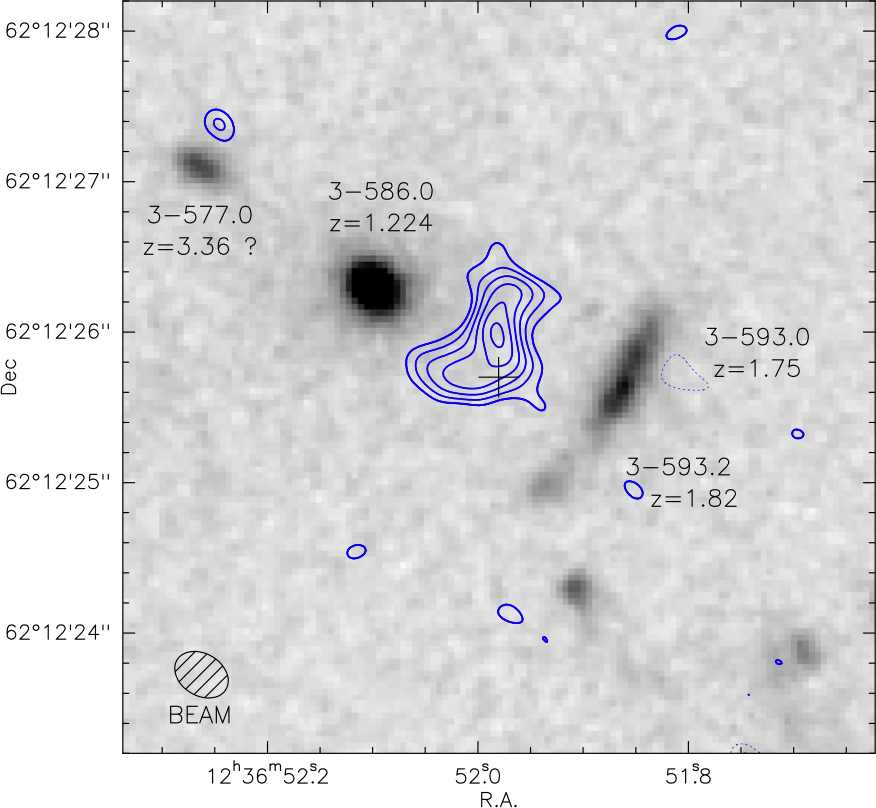}
\includegraphics[angle=-0,width=8.5cm]{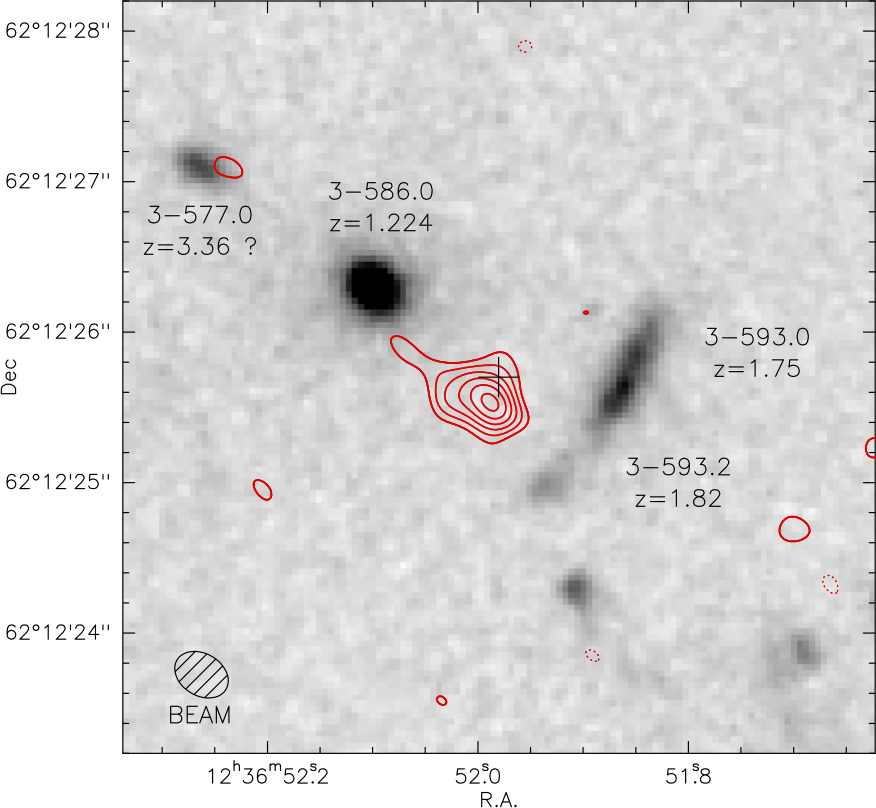}
\caption[Maps of blue- and redshifted C$^+$]
{Maps of blueshifted ({\it left}) and redshifted ({\it right}) C$^+$
emission in HDF\,850.1, with the 0.98\,mm continuum subtracted.  These
two channels are each 470\,\kms\ wide, centered on $-$205\,\kms\ and
+265\,\kms, with integration ranges of $-440$\,km\,s$^{-1}$ to
$+30$\,km\,s$^{-1}$ ({\it left}) and $+30$\,km\,s$^{-1}$ to
$+500$\,km\,s$^{-1}$ ({\it right}). These velocities are relative to
307.267\,GHz (C$^+$\,158\,$\mu$m at $z=5.1853$).  Contours are
$-3\,\sigma$ (dashed), $+3\,\sigma$, and then go up in steps of 1\,$\sigma$ =
0.20\,Jy\,beam$^{-1}$\,\kms . The blueshifted, C$^+$ North peak ({\it
left}) is 1.6\,Jy\,beam$^{-1}$\,\kms , and its spatially integrated
flux is 6.9\,Jy\,\kms .  The redshifted, C$^+$ South peak ({\it
right}) is 1.4\,Jy\,beam$^{-1}$\,\kms , and its spatially integrated
flux is 3.0\,Jy\,\kms .  The beam (lower left, in each panel), is
0.38$''\times0.29''$ at PA$=59^\circ$ (FWHM), with $T_b^\prime/S =
118$\,K\,Jy$^{-1}$. The cross marks the phase reference position, at
12:36:51.980, +62:12:25.70 (J2000). The C$^+$-line contours are
superposed on a greyscale version (Downes et al.\ 1999) of the {\it
BVI} image from the Hubble Deep Field.
}
\label{blue-red maps}
\end{figure*}

\begin{figure*} \centering
\includegraphics[angle=-0,width=17.0cm]{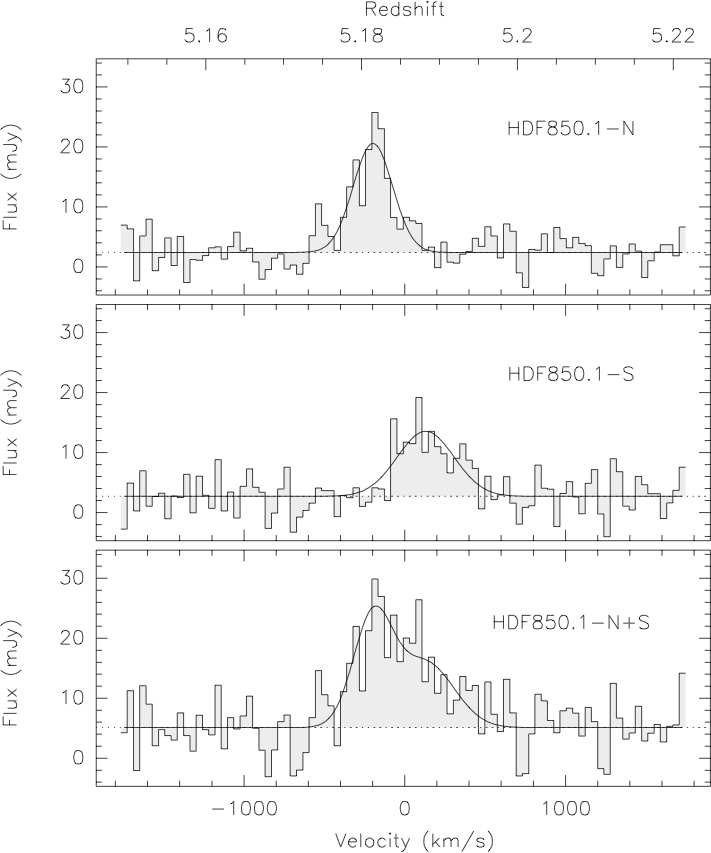}
\caption[C$^+$ Spectra]{
Spatially-integrated spectra of the C$^+$ $^2$P$_{3/2}\rightarrow
^2$P$_{1/2}$ line, from HDF\,850.1-North and South (top and middle
spectra), and integrated over the entire source (bottom spectrum).
The lines are shown above the dust continuum, which is 2.2\,mJy for
the North galaxy (top spectrum), 2.4\,mJy for the South galaxy
(middle spectrum), and 4.6\,mJy for the whole source (bottom
spectrum).  Channel widths are 39\,km\,s$^{-1}$, and velocities are
relative to 307.267\,GHz (C$^+$\,158\,$\mu$m at $z=5.1853$).  The
solid curves are Gaussian fits to the line profiles, with a
double-Gaussian fit for the entire source (bottom spectrum). The
upper horizontal scale shows the range of cosmological redshifts
covered by our observing bandwidth.  We interpret the spread in line
velocities as being due to kinematic motions (lower horizontal scale,
in \kms), not different cosmological redshifts.  (These spectra are
available as ascii tables at the CDS via anonymous ftp to cdsarc.u-strasbg.fr.)
}
\label{h850-C+spectra}
\end{figure*}

\section{New long-baseline observations and results}
To further investigate HDF\,850.1's morphology, we re-observed its C$^+$ line
and adjacent millimeter continuum with the IRAM Plateau de Bure
interferometer in the A configuration, with long baselines to 760\,m,
in February 2012, and in the B configuration in March 2012. We
combined the new data with the earlier observations (Walter et
al.\ 2012) from November 2011, in the shorter-baseline C-configuration.
All data were taken in excellent radio ``seeing'' (phase errors
corresponding to 0.1$''-$0.3$''$) and low zenith opacity (0.13$-$0.35
at 0.98\,mm wavelength).  The observations were made at 307.267\,GHz
(0.98\,mm), the frequency of the redshifted $^2$P$_{3/2}\rightarrow
^2$P$_{1/2}$ fine-structure line of C$^+$ at a redshift of
$z=5.1853$. Receiver temperatures were 45 to 65\,K.  The wideband
correlator covered the 305.6$-$309.2\,GHz frequency band in both
polarizations. We calibrated the bandpass on 3C84 and
0851$+$202. Time-dependent gains were calibrated in amplitude and
phase by interleaving observations of HDF\,850.1 with 1150+497 and
1300+580. The flux density scale is relative to an assumed flux
density of 2.35\,Jy of MWC349 at 307\,GHz. The uncertainties in the
flux calibration are less than 10\%.  For maximum sensitivity in line
and continuum, we used natural weighting for the visibilities. The
projected baselines, ranging from 23 to 780\,k$\lambda$, gave a
synthesized beam of 0.38$''\times $0.29$''$ at PA +59$^\circ$.  For
the nominal phase center, we used the HDF\,850.1 position measured by
Downes \etal\ (1999; 12:36:51.980, $+$62:12:25.70, J2000).  For our new
measurements, the positional uncertainty of this nominal phase center
was verified on 1150+497 and 1300+580, and estimated to be smaller
than 0.03$''$, or one-tenth of the synthesized beam.

For the continuum, we used two line-free regions adjacent to the C$^+$
line, covering a total of 2\,GHz. For the C$^+$ line, we subtracted
the continuum in the (u,v) plane, and binned the data into 40\,MHz
(39\,\kms ) channels. The r.m.s.\ noise in the maps is
0.21\,mJy\,beam$^{-1}$ in the continuum, and 1.42\,mJy\,beam$^{-1}$
per 40\,MHz channel in the line.

Figure~1 shows two velocity-integrated maps of the C$^+$ line,
with the continuum subtracted.   They clearly show
two distinct peaks, with a separation of 0.3$''$, or a projected distance of
2\,kpc,\footnote{We assume $H_0$=71\,km\,s$^{-1}$\,Mpc$^{-1}$, $\Omega_M$=0.27, and $\Omega_{\Lambda}$=0.73, which gives an angular scale of  6.295\,kpc (arcsec)$^{-1}$ at $z$ = 5.185.}
which we interpret as evidence for a merger of two galaxies,
with the blueshifted galaxy to the north, and the redshifted galaxy to
the south.  The lower-intensity emission around the two peaks covers
an overall extent, down to the $3\,\sigma$ contours, of 1.1$''$ (projected size 7\,kpc), in the southeast-to-northwest direction.  Table~1 lists the source
positions, fluxes, and apparent sizes derived from (u,v) fits to these maps.  We
see no ordered pattern of velocities suggesting rotation, neither in
channel maps, nor in second-moment velocity contour maps (not shown
here).  This absence of an ordered, monotonically progressing velocity
pattern is similar to that observed in the sample of submillimeter
galaxy (SMG) mergers studied by Men\'endez-Delmestre et al.\ (2013).

Further evidence for a merger comes from the spectral profiles of
the two components.  Figure~2 shows the C$^+$ spectra, spatially
integrated over the blueshifted North galaxy, the redshifted
South galaxy, and the total spectrum, spatially integrated
over both North+South galaxies together. Both the North and South C$^+$
line
profiles appear roughly Gaussian, with large linewidths,  $\sim 300$
and  $\sim 410$\,\kms , respectively.  These are linewidths
typical of individual galaxies, or individual, rapidly rotating, compact
circumnuclear disks, rather than linewidths expected at opposite ends of
 a large ($\sim 10$\,kpc), quiescent, rotating disk.

Furthermore, the blueshifted, North galaxy (Fig.~1, left) is
spatially extended, while its line profile (Fig.~2, upper) has a
somewhat smaller width ($\sim 300$\,\kms ).  In contrast, the
redshifted, South galaxy is compact (Fig.~1, right), less
intense, and wider in velocity ($\sim 410$\,\kms ).

From these spatial and spectral differences,
and the large total width of the line (940\,\kms ),
we speculate that the two components may not be in the same
plane (so not in a disk), and that we are looking at the merging of
two galaxies with different inclinations,  with the North
galaxy possibly more face-on than the South galaxy.

Figure~3 (left) is a map of the C$^+$ line flux density, in
mJy\,beam$^{-1}$, averaged over the line full width of 940\,\kms. The
contours have an irregular V-like shape, similar to those of
interacting systems, rather than a single, undisturbed disk galaxy.
Figure~3 (middle) shows a map of the redshifted 158\,$\mu$m dust continuum
emission, and Fig.~3 (right) shows
the ratio of the velocity-integrated C$^+$ emission to
the redshifted 158\,$\mu$m continuum over the source. The continuum
map looks slightly different from the line maps in Fig.~1,
probably because of the difference in signal-to-noise ratios between
the line and continuum maps.  Furthermore,
the broad-band continuum is a blend of dust emission from
both the North and South galaxies,
in the region where these two components overlap.
The line-to-158\,$\mu$m continuum ratio (Fig.~3, right) is
nearly constant across the source, with a value of about 1.5.

\begin{table*}
\caption{Positions, sizes, and fluxes of the HDF\,850.1 merger galaxies.}
\begin{tabular}{lll ccc ccc c}
\hline
 &R.A. &Dec.	&Major &Minor	&P.A.    &C$^+$  &Velocity   &C$^+$   &Dust	
\\
 &12$^{\rm h}$36$^{\rm m}$
&62$^\circ 12'$  &axis &axis     &        &Peak   &at peak   &flux  &flux
\\
Data
&J2000  &J2000 &arcsec &arcsec  &deg.    &mJy    &\kms      &Jy\,\kms &mJy
\\	
\hline
\multicolumn{5}{l}{ {\bf HDF\,850.1 C$^+$ line:} }\\
\multicolumn{5}{l}{North galaxy}
\\
$-$440 to $+$30\,\kms:
	&51.989$^{\rm s}$
	&$25.88''$
        &$0.8''$
        &$0.4''$
	&$-46^\circ$
        &18.8
        &$-$200
	&6.9
        &2.2
\\
\multicolumn{5}{l}{ South galaxy}
\\
+30 to +500\,\kms :
	&51.993$^{\rm s}$
	&$25.57''$
	&$0.4''$
	&$0.4''$
	&---
        &11
        &+130
	&3.0
        &2.4
\\
\\
\multicolumn{5}{l}{ {\bf HDF\,850.1 dust continuum at 0.98\,mm:} }\\
North+South galaxies
\\
together:
        &51.993$^{\rm s}$
	&$25.70''$
	&$0.9''$
	&$0.3''$
	&$-23^\circ$
	&---
        &---
        &9.9
        &4.6
\\
\multicolumn{5}{l}{ {\bf Lensing elliptical galaxy:} }\\
3-586.0 at $z$ = 1.224:
        &52.090$^{\rm s}$
        &26.30$''$
        &---
        &---
        &---
        &---
        &---
        &---
        &$<0.9$
\\
\hline
\multicolumn{10}{l}{{\bf Notes:}  C$^+$ and dust positions, sizes, and fluxes are from the maps and from Gaussian fits in the (u,v) plane.}
\\
\multicolumn{10}{l}{C$^+$ peak fluxes and velocity peaks are from Gaussian fits to the spatially-integrated spectrum.}
\\
\multicolumn{10}{l}{Errors: positions:
 $\pm 0.004^{\rm s}$ in R.A.\ and $\pm 0.03''$ in Dec.;
 sizes: $\pm 0.1''$; P.A.: $\pm 5^\circ$; C$^+$ fluxes: $\pm$10\%;
dust flux $\pm$20\%.}
\\
\multicolumn{10}{l}{
Velocities of the C$^+$ line peaks are relative to
307.267\,GHz (C$^+$\,158\,$\mu$m at $z=5.1853$). Errors are $\pm$20\,\kms .}
\\
\multicolumn{10}{l}{Position of the elliptical galaxy is
from Barger et al.\ (2008).}
\\
\multicolumn{10}{l}{Dust continuum limit for the elliptical is
the 5\,$\sigma$ limit from this paper, Fig.~3 (middle).}
\\
\end{tabular}
\end{table*}

\begin{table*}
\caption{Derived quantities for the HDF\,850.1 merger galaxies, and the effect of lensing.}
\begin{tabular}{l ccc }
\hline
 Parameter  &North 	 &South          &Dust \\
            &galaxy      &galaxy     &continuum,     \\
            &C$^+$ line  &C$^+$ line &both galaxies \\
\hline
\multicolumn{4}{l}{ {\bf Uncorrected for lensing:} }\\
Measured ($\Delta\alpha,\Delta\delta$) offset from elliptical
	&($-0.71'',-0.40''$)
	&($-0.68'',-0.71''$)
    &($-0.68'',-0.58''$)
\\
Apparent luminosity [ \Lsun ]
	&$L_{\rm C^+} = 2.8  \times 10^9$
	&$L_{\rm C^+} = 1.2  \times 10^9$
	&$L_{\rm FIR} = 6.0  \times 10^{12}$
\\
Apparent line luminosity [ \Kkmspc ]
        &$L^\prime_{\rm C^+} = 3.2  \times 10^{10}$
	&$L^\prime_{\rm C^+} = 1.4  \times 10^{10}$
        &---
\\
\\
\multicolumn{4}{l}{ {\bf Corrected for lensing:} }\\
True, undeflected ($\Delta\alpha,\Delta\delta$) offset from elliptical
        &($-0.45'',-0.25''$)
	&($-0.50'',-0.50''$)
	&($-0.47'',-0.43''$)

\\
lens magnification factor
        &1.7
        &1.5
        &1.6
\\
Corrected luminosity [ \Lsun ]
	&$L_{\rm C^+} = 1.6  \times 10^9$
	&$L_{\rm C^+} = 8.2  \times 10^8$
	&$L_{\rm FIR} = 3.8  \times 10^{12}$
\\
Corrected line luminosity [ \Kkmspc ]
        &$L^\prime_{\rm C^+} = 1.9  \times 10^{10}$
	&$L^\prime_{\rm C^+} = 9.4  \times 10^{9}$
        &---
\\
Gas Mass (H$_2$+He)  [\Msun ]
        &$M_{\rm gas} = 1.5  \times 10^{10}$
	&$M_{\rm gas} = 7.5  \times 10^{9}$
        &---
\\
Corrected angular diameters
	&$~0.4''$
	&$~0.3''$
	&$~0.4''$
\\
True radius
        &1.3\,kpc
        &0.9\,kpc
        &1.3\,kpc
\\
\hline
\multicolumn{4}{l}{{\bf Notes:} Positions are from Gaussian fits in the (u,v) plane, with errors
 of $\pm 0.004^{\rm s}$ in R.A.\ and $\pm 0.03''$ in Dec.}
\\
\multicolumn{4}{l}{$L_{\rm FIR}$ is for emitted wavelengths 40 to 120\,$\mu$m;
$L_{\rm IR}$ (5 to 1000\,$\mu$m) is typically 30\% higher.}
\\
\multicolumn{4}{l}{Lens model: singular isothermal sphere at $z$ = 1.224, with velocity dispersion $\sigma$ = 150\,\kms .}\\
\multicolumn{4}{l}{Derived values are for $D_A$ = 1.299\,Gpc at $z$ = 5.185,
and 6.295\,kpc\,(arcsec)$^{-1}$,  for assumed values of}\\
\multicolumn{4}{l}{$\Omega_M$ = 0.27, $\Omega_{\Lambda}$ = 0.73, and $H_0$ = 71\,\kms\,Mpc$^{-1}$ (for formula, see e.g., Buchalter et al.\ 1998).}
\\
\end{tabular}
\end{table*}

\section{Lensing by the elliptical galaxy 3-586.0 at $z$ = 1.224}
The C$^+$-line contours in Fig.~1 are superposed on a greyscale
version (Downes et al.\ 1999) of the {\it BVI} image from the Hubble
Deep Field, showing, as has been known for some time, that there are
no optical counterparts at the positions of the merger galaxies in
HDF\,850.1.  This superposition also shows clearly that the lines of
sight to both of the merger components of HDF\,850.1 pass close to the
compact elliptical galaxy 3-586.0 (Williams et al.\ 1996), so both of
the merger galaxies are subject to weak gravitational lensing.

With our new measured positions for the HDF\,850.1 North and South
galaxies, we re-computed the gravitational lensing effect by the $z$ =
1.2 elliptical galaxy 3-586.0.  We took the position of this
elliptical galaxy from Barger et al.\ (2008).
Our measured C$^+$ position of the North galaxy is at
($\Delta\alpha, \Delta\delta$) = ($-0.71''$, $-0.40''$) from the
center of the elliptical; the South galaxy is at ($-0.68''$,
$-0.71''$).  We modeled the lens as a singular isothermal sphere, at
redshift $z$ = 1.224 (Barger et al.\ 2008), with velocity dispersion
150\,\kms\ (Dunlop et al.\ 2004). We assumed reasonable values for the
true (undeflected) offsets of the two HDF\,850.1 galaxies in R.A.\ and Dec.,
and slowly varied these assumed offsets until the
predicted lensed images of the two galaxies fell at their observed (deflected)
displacements from the elliptical.  Table~2 gives the results.

In this model, the lensing magnification is 1.7 for the North galaxy,
and 1.5 for the South galaxy.  The predicted galaxy images are
arcs with long axes roughly along
PA = $-45^\circ$ (southeast to northwest), and this is approximately
the mean direction of the observed C$^+$ intensity contours.
When the observed image diameters listed in
Table~1 are corrected for these magnifications, then the corrected
image sizes become nearly circular, with radii of 1.3 and 0.9\,kpc
for the C$^+$ emitting regions
in the North and South galaxies, respectively.

\begin{figure*} \centering
\includegraphics[angle=-0,width=5.9cm]{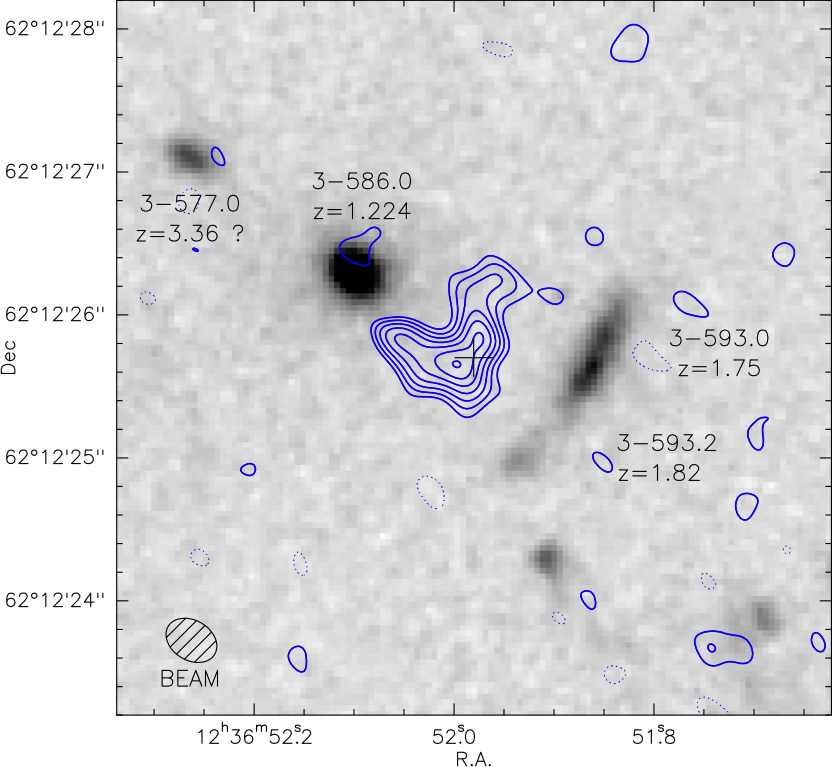}
\includegraphics[angle=-0,width=5.9cm]{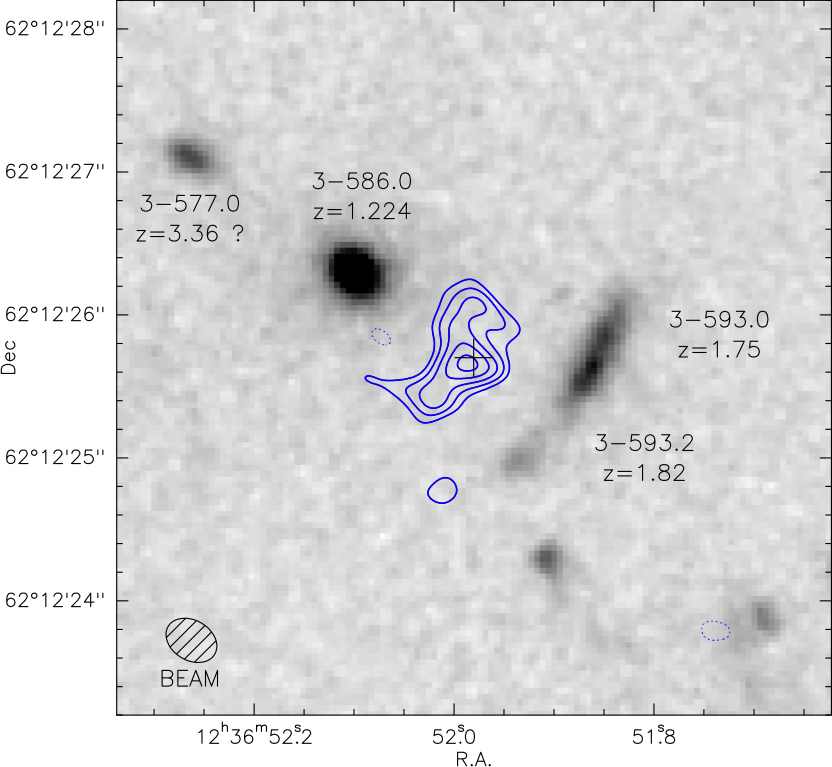}
\includegraphics[angle=-0,width=5.9cm]{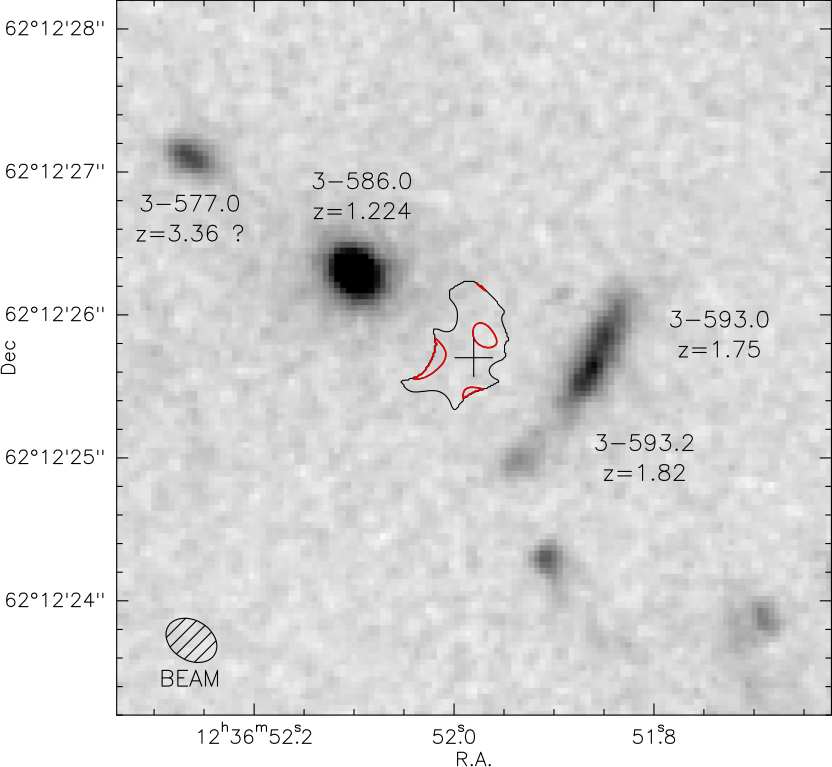}
\caption[Line, Continuum and L/C Ratio]{
Maps of the velocity-averaged C$^+$ flux density {\it (left)}, the
0.98\,mm continuum {\it (middle)} and the line-to-continuum ratio
{\it (right)}.  The synthesized beam is 0.38$''\times 0.29''$ (lower
left of each panel). In these maps, the cross marking the phase
reference position, and the background greyscale image, are the same
as in Fig.~1.

{\it Left panel:} The C$^+$ line flux density, in mJy\,beam$^{-1}$,
is averaged over the line full width of 940\,\kms. Line contours are
$-3\,\sigma$ (dashed), $+3\,\sigma$, and then going up in steps of
1\,$\sigma = 0.24$\,mJy\,beam$^{-1}$. The peak is
2.2\,mJy\,beam$^{-1}$, and the spatial integral is 9.9\,mJy.

{\it Middle panel:} The dust continuum was integrated over 2\,GHz of
line-free channels. Continuum contours are $-3\,\sigma$ (dashed),
$+3\,\sigma$, and then going up in steps of 1\,$\sigma =
0.19$\,mJy\,beam$^{-1}$.  The peak continuum emission is
1.4\,mJy\,beam$^{-1}$.  The spatially-integrated continuum flux
density is 4.6\,mJy.

{\it Right panel:} The line-to-continuum ratio is roughly constant
across the map; contour values are 1.0 (black) and 2.0 (red).  In
these units, the line-to-continuum ratio is dimensionless, and its
typical value across the source is 1.5.
}
\label{line, continuum+line-to-cont-ratio}
\end{figure*}

\section{Dust brightness temperature and dust optical depth at 158\,$\mu$m}
\subsection{Measured brightness temperatures}
In radio astronomy, we measure brightness temperatures, which are
defined by the Rayleigh-Jeans formula (e.g., Pawsey \&
Bracewell 1955). This definition has nothing to do with whether radio
sources are ``in the Rayleigh-Jeans regime'', or even whether the
radiation is thermal or nonthermal.  Defined this way, the measured
brightness temperatures are extremely useful because they are directly
proportional to source brightness in units of Jy per solid angle.  We
can therefore write radiative transfer equations directly in terms
of our measured brightness temperatures, in a much less complicated
form than if we would use thermodynamic quantities like dust grain
temperatures, spectral line excitation temperatures, or the 2.7\,K
cosmic background radiation temperature.
Fortunately, it is easy to
convert from measured brightness temperature to the radiation
temperature of an equivalent blackbody, by equating the Rayleigh-Jeans
formula to the Planck formula.  For, e.g.,  the cosmic background,
this yields:
\begin{equation}
T^\prime_{bg} = (h\nu /k)(\exp (h\nu /kT_{bg}) -1)^{-1}
\end{equation}
where $T^\prime_{bg}$ is the measured brightness temperature
at the telescope, and $T_{bg}$ is the thermodynamic (Planck) temperature.
For example, at the frequency of this
project, 307\,GHz, if we observed the 2.7\,K cosmic background, then at the
telescope, we would measure a brightness temperature of only 0.063\,K,
not the 2.7\,K radiation temperature of the cosmic background.

We now show that because the measured brightness temperature of the
HDF\,850.1 continuum source is relatively high, the dust radiation
must be nearly optically thick.

\subsection{Application to the dust continuum of HDF\,850.1}
We measure a dust continuum flux density of 4.6\,mJy at 0.98\,mm, and
a source diameter of 0.4$''$.  These parameters correspond to a
measured brightness temperature of $T_b^\prime$ = 0.54\,K at 0.98\,mm.
 Corrected for (1+$z$), this corresponds
to $T_b^\prime$ = 3.3\,K at 158\,$\mu$m.
Here, the primed symbol $T_b^\prime$ indicates the standard radio
astronomy brightness temperatures, defined via the Rayleigh-Jeans formula,
which is directly proportional to the brightness, in units
of Jy\,sr$^{-1}$.

The gravitational lensing has no effect here, because the lensing
amplification of the source flux and the lensing magnification of the
source area are the same, so the gravitational lensing preserves
brightness temperature.

For simple radiative transfer in a homogeneous medium, the observed
brightness temperature is

\begin{equation}
 T_b^\prime = (T_d^\prime - T_{bg}^\prime)(1-e^{-\tau_d})
\end{equation}
where $T^\prime_d$ and $T^\prime_{bg}$ are the Rayleigh-Jeans equivalents of
the dust and background temperatures, and $\tau_d$ is the dust optical depth.

At $z$ = 5.2, the cosmic background (Planck) temperature is 16.9\,K, and
at 158\,$\mu$m, its Rayleigh-Jeans equivalent is $T^\prime_{bg}$ =
0.4\,K, so at this wavelength, even at a redshift of 5.2, the
radiative-transfer correction for the cosmic background is small.

This means that if $\tau_d$ were $\gg 1$, the dust temperature would
be $T_d^\prime = T_b^\prime$ + 0.4\,K, or 3.7\,K, which corresponds to
a Planck temperature of $T_d$ = 28\,K.  This is the minimum dust
temperature allowed by the observations. It corresponds
to the limit of the dust being highly optically thick at 158\,$\mu$m.
In general, in eq(2),
for a measured brightness temperature,  $T_b^\prime$,
the inferred dust temperature, $T_d^\prime$, varies inversely as the
factor containing the optical depth. If, for example $\tau_d$ = 0.6,
then  $T_d^\prime$ increases to 7.7\,K, which corresponds to
a Planck temperature of $T_d$ = 36\,K.

We can also reverse the argument and adopt a dust temperature to
derive the dust optical depth.  Fits to the measured continuum fluxes
of HDF\,850.1 with ULIRG templates (e.g., Walter et al.\ 2012)
indicate that the dust temperature is 45\,K,
or an equivalent Rayleigh-Jeans dust temperature of 14\,K at
158$\mu$m.  From eq(2), for our measured continuum brightness
temperature, this gives a dust optical depth of $\tau_d$ =
0.4 at 158\,$\mu$m, so the dust is starting to become optically thick as
one approaches the peak of the FIR continuum.  If the true dust temperature
is somewhat lower, like the 30 to 35\,K inferred
for many submillimeter galaxies,
then the dust optical depth will be close to unity at 158\,$\mu$m.

\section{Excitation temperature of the C$^+$ line}
Similar reasoning holds for the C$^+$ line.  From the observed
line fluxes and source angular sizes (Tables 1 and 2), we obtain peak
brightness temperatures, corrected by $(1+z)$, of 10.7 and 4.7\,K for
HDF\,850.1 N and S respectively, if the sources have a Gaussian shape,
and 21.4 and 7.9\,K respectively, if they are disks.  From the analogs
of eqs(1) and (2) for line excitation temperature, we can show that
these brightness temperatures correspond to line excitation
temperatures in the range 30 to 50\,K if the C$^+$ line is optically
thick ($\tau \gg 1$).  This is exactly in the range deduced for the
dust temperatures from the SED fitting by Walter et al.\ (2012), and
for the gas kinetic temperatures that are consistent with the
escape-probability fits by Walter et al.\ to their observed lines in
the CO ladder.  Because we are directly observing C$^+$ line
brightness temperatures comparable to the gas and dust temperatures,
we conclude that the C$^+$ lines are optically thick.

A second argument that the C$^+$ lines are optically thick comes from
the fact that, in brightness-temperature units (\Kkmspc ), the C$^+$
and CO(1--0) line luminosities are roughly the same (see discussion in
Section~8).  Because the C$^+$ and CO line profiles are very similar,
with nearly the same linewidths (Walter et al. 2012), then if the
emitting areas are about the same, the C$^+$ and CO line brightness
temperatures must also be roughly the same.  Because, in general, the CO(1--0)
lines are optically thick and trace the gas temperature, the same must
then also be true of the C$^+$ line.

A third argument that C$^+$ is optically thick
comes from the escape-probability
program RADEX (van der Tak et al.\ 2007).
For $T_{\rm kin} \sim 30$\,K and for plausible
H$_2$ densities $> 10^3$\,cm$^{-3}$ and
C$^+$ column densities $>10^{18}$\,cm$^{-2}$, the program predicts a
C$^+$ line excitation temperature of 29\,K,
about the same as the gas kinetic temperature, and a
line optical depth $\tau_L >$ 6.

Similar conclusions about the optical depth of the C$^+$ 158\,$\mu$m line
in galaxies have been reached by many other authors (e.g., Crawford et al.\ 1985;
Stacey et al.\ 1991a; Stacey et al.\ 1991b (from observations of  $^{13}$C$^+$);
Mashian et al.\ 2013).

Note also that although the energy needed to singly ionize carbon is 11\,eV, or $E/k
\sim 10^5$\,K, the excitation temperature of the C$^+$ 158\,$\mu$m
line is much lower,  30 to 50\,K.  This excitation
may be provided by collisions with hydrogen molecules at the kinetic
temperatures typical of warm molecular clouds in starburst galaxies.
We therefore assume the C$^+$ line excitation temperature
 is the same as the dust and gas kinetic temperatures.

If all this is correct, then the C$^+$ line must be opaque,
because with an excitation temperature in this ``low'' range, an
optically thin C$^+$ line would be hard to
detect at cosmological distances.

We can now use these ideas to predict
the C$^+$-to-dust continuum flux ratio. We apply the same argument as
used by Downes et al. (1993) to predict the ratio of the CO line flux
to 100\,$\mu$m continuum flux in ULIRGs.

In this case, however, we are trying to predict the C$^+$-to-dust flux ratio {
\it at the same rest wavelength}, namely 158\,$\mu$m, so
to first order, the C$^+$-to-dust continuum flux ratio simplifies to
\begin{equation}
{S_L\over S_C} = { {(T_{ex}^\prime - T_{bg}^\prime)(1-e^{-\tau_L}) }
\over {(T_d^\prime - T_{bg}^\prime)(1-e^{-\tau_d})} }
{\Omega_L \over \Omega_C}
\end{equation}
where $T^\prime_{ex}$, $T^\prime_{bg}$, and $T^\prime_d$
are the Rayleigh-Jeans equivalents of the line
excitation temperature, the cosmic background temperature,
and the dust temperature.
The optical depths are $\tau_L$ for the line, and $\tau_d$ for the dust.
We assume that the solid angles, $\Omega_L$ and
$\Omega_C$, of the C$^+$ line and dust continuum sources are the same.

So for an optically thick C$^+$ line, with $\tau_L \gg 1$, and with
similar line and continuum source sizes, $\Omega_L \approx \Omega_C$, and
similar line excitation and
dust temperatures, $T_{ex}^\prime \approx T_d^\prime$, eq(3) becomes
\begin{equation}
{S_L\over S_C} = { 1
\over {1-e^{-\tau_d}} }
\end{equation}
or, if the dust is optically thin, simply
\begin{equation}
{S_L\over S_C} = { 1
\over \tau_d }
\end{equation}
For a dust optical depth close to unity at the FIR
peak near 100\,$\mu$m, as is the case for many ULIRGs and dusty,
high-redshift submillimeter galaxies, the line-to-continuum flux
ratio should approach unity.

For HDF\,850.1, we can compare these ideas directly with the
observations.  On the velocity-integrated map in Fig.~3, left,
the line flux, integrated over the region shown on the
continuum map, is  6.5\,Jy\,\kms .
We can divide this integrated line flux by the full linewidth
to zero power, or 940\,\kms , to obtain an average flux density over the
line, of 6.9\,mJy.
Dividing this average line flux density by the spatially-integrated dust
flux density of 4.6\,mJy gives a line-to-continuum ratio of 1.5, as
shown in Fig.~3 ({\it right}).
This line-to-continuum ratio of about 1.5 would fit
eq(4) if the dust optical depth were about unity at 158\,$\mu$m.
This dust opacity is even higher than we estimated in section~4
by considering the continuum brightness temperature alone,
so our conclusion is the same.
The observed line and continuum brightness temperatures
imply high C$^+$ line optical depth, ``low'' C$^+$ excitation
temperature, of the same order as the dust temperature, with the dust
starting to become optically thick near 158\,$\mu$m.

\section{Comparison of the line-to-continuum ratio in other high-$z$ galaxies}
So far, there are only 4 high-$z$ objects with subarcsec-beam
measurements of the C$^+$ line and the
158\,$\mu$m dust continuum (HDF\,850.1, BR1202$-$07, J1148+5251, 0952$-$0115).
Usually the integrated C$^+$
line luminosity is compared with the far-IR or entire infrared
luminosity, to see if the C$^+$ line is correlated with the star
formation rate.  In this paper, our goal is different: we
explore the physics of the C$^+$ line, and estimate its optical
depth and likely excitation temperature.

Table~3 lists some of the measurements in the literature.
Although the argument in the previous section can be given
in terms of the peak line brightness temperature, for consistency with
our Fig.~3, we prefer comparing HDF\,850.1 with other high-$z$
galaxies in terms of mean brightness temperature (or flux density per
beam area), averaged over the entire line width, i.e., full width to
zero power.

Our comparison shows that unlike the ratios of integrated C$^+$ line
luminosity to far-IR luminosity, which have a wide scatter, the
observed ratios of mean line brightness temperature to 158\,$\mu$m
dust brightness temperature are mostly about 2-to-1, within a range of a
factor of two (Table~3 and Fig.~4).
Furthermore, unlike the C$^+$-to-FIR
luminosity ratios, which seem to have a strong separation between AGNs and
starbursts (Maiolino et al.  2009; Graci\'a-Carpio et al.\ 2011), the
C$^+$-to-158\,$\mu$m brightness temperature ratios (or flux density
ratios) only show at most
a factor-of-two difference between these two classes of
sources.
Our mean ratio for the submillimeter galaxy starbursts is
$1.1\pm 0.3$, versus a ratio of $2.0\pm1.2$ for the quasar host galaxies,
so their mean values actually agree, within the error bars.

This nearly constant ratio confirms that the C$^+$ line is optically
thick, with an excitation temperature roughly equal to the dust temperature,
and that the dust itself is also nearly optically thick (dust optical
depth in the range 0.1 to 1), and that the small variations in the
ratio from source to source can mostly be explained by small
variations in the dust opacity.  To first order, we may say
that as the dust radiation starts to become optically thick, it no longer
traces mass, as it does at millimeter wavelengths, where the dust is
optically thin, but instead it starts to trace the dust brightness
temperature at the $\tau$ = 1 surface.  Similarly, the optically thick
C$^+$ line (like the CO(1--0) line) traces the gas brightness temperature.
Insofar as the gas and dust brightness temperatures are about the same,
on kpc scales, in the high-$z$ sample in Table~3, the C$^+$-to-158\,$\mu$m-continuum brightness temperature ratio will be roughly constant.

In contrast, the C$^+$-to-FIR luminosity ratios vary by at least a factor of
100, as a function of $L_{\rm FIR}$ (see, e.g., Cox et al.\ 2011, their fig.6;
Carilli \& Walter 2013, their fig.6).
The much larger variations in the C$^+$-to-FIR luminosity ratios, are due
not only to the variations in dust opacity, but also to the fact that the
C$^+$ luminosity varies only as $T^\prime_b \Delta V$, i.e., Rayleigh-Jeans
brightness temperature, whereas for dense dust sources, the far-IR luminosity
varies as (Planck) $T_d^4$.
This means that the C$^+$-to-FIR luminosity
ratios have a much greater sensitivity to dust temperature variations
than do the  C$^+$-to-158\,$\mu$m-continuum flux density ratios.
In particular, in ``warm'' (AGN) sources, the C$^+$-to-FIR luminosity
ratios will decrease, relative to ``cool'' sources, not because the
C$^+$ is somehow ``quenched'',
but simply because the FIR luminosity of the warm dust is so large.

\begin{table*}
\caption{C$^+$ line and 158\,$\mu$m dust continuum parameters, for
HDF\,850.1 and other high-redshift galaxies.}
\begin{tabular}{lccc cccl }
\hline
 Source
&         &Spatially-  &C$^+$     &C$^+$ line  &158\,$\mu$m   &Line to\\
&Redshift &integrated  &linewidth &average flux   &continuum     &dust  & Ref.\\
&         &line flux   &$\Delta V$  &density$^{\rm a}$
                                               &flux density  &flux density \\
&(z)      &(Jy\,\kms ) &(\kms )   &(mJy)       &(mJy)         &ratio     \\
\hline
\multicolumn{8}{l}{ {\bf High-redshift submillimeter galaxies (SMGs):} }\\
{\bf HDF\,850.1}
     &5.2     &6.5$^{\rm b}$      &470         &6.9      &4.6   &1.5    &1 \\
HLS0918      &5.2     &107       &700         &76       &103   &0.7    &2 \\
BR1202$-$07 NW
             &4.7     &15        &425         &17.7     &19.0  &0.9    &3, 4 \\
J1424+0223   &4.2     &107       &690         &77.5     &90    &0.8    &5   \\
HFLS3        &6.3     &14        &470         &14.9     &13.9  &1.1    &6  \\
ALESS 61.1   &4.4     &2.5       &230         &5.4      &4.3   &1.3    &7 \\
ALESS 65.1   &4.4     &5.4       &470         &5.7      &4.2   &1.4    &7 \\ \\
\hline
\multicolumn{6}{l}{ {\bf Mean C$^+$ line-to-dust ratio for SMGs:} }&{\bf $1.1\pm0.3$}
\\ \\
\multicolumn{8}{l}{ {\bf High-redshift quasar hosts (QSOs):} }\\
BR1202$-$07 SE &4.7   &15        &600        &12.5    &18    &0.7      &3, 4 \\
J0129$-$0035 &5.8     &2.0       &194        &5.2     &2.6   &2.0      &8 \\
J0210$-$0456 &6.4     &0.27      &300        &0.45    &0.12  &3.7      &9 \\
J0952$-$0115 &4.4     &20.1      &193        &52.1    &15.6  &3.3      &10 \\
J1044$-$0125 &5.8     &1.7       &420        &2.0     &3.1   &0.7      &8 \\
J1120+0641   &7.1     &1.0       &235        &2.2     &0.61  &3.6      &11  \\
J1148+5251   &6.4     &4.1       &350        &5.9     &4.5   &1.3     &12, 13, 14 \\
J1319+0950   &6.1     &4.3       &518        &4.1     &5.2   &0.8      &8    \\
J2054$-$0005 &6.0     &3.5       &242        &7.1     &3.1   &2.3      &8  \\
J2310+1855   &6.0     &9.0       &392        &11.4    &9.1   &1.3      &8    \\
\\
\hline
\multicolumn{6}{l}{ {\bf Mean C$^+$ line-to-dust ratio for QSOs:} }&{\bf $2.0\pm1.2$}
\\
\\
\multicolumn{8}{l}{{\bf Notes:} $^{\rm (a)}$ C$^+$ average line flux density = $\int SdV / 2\Delta V$.}
\\
\multicolumn{8}{l}{$^{\rm (b)}$ In the region where the C$^+$ and continuum maps overlap.}
\\
\multicolumn{8}{l}{References: (1) this paper, (2) Rawle et al.\ 2013,
(3) Wagg et al.\ 2012, (4) Carilli et al.\ 2013, (5) Cox et al.\ 2011,}
\\
\multicolumn{8}{l}{(6) Riechers et al.\ 2013, (7) Swinbank et al.\ 2012,
(8) Wang et al.\ 2013,  (9) Willot et al.\ 2013,}
\\
\multicolumn{8}{l}{ (10) Gallerani et al.\ 2012, (11) Venemans et al.\ 2012, (12) Maiolino et al.\ 2005,}
\\
\multicolumn{8}{l}{(13) Maiolino et al.\ 2012, (14) Walter et al.\ 2009.}
\\
\end{tabular}
\end{table*}

\section{Implications for the C$^+$ line physics}
In normal, low-redshift quiescent galaxies like the Milky Way, the
C$^+$ line comes from a mixture of different environments: on 1-pc
scales, from the widely-studied photon-dominated regions (PDRs) at
cloud boundaries, on larger scales, from the neutral gas, and from
within the bulk of the molecular clouds themselves (see Pineda et al.\
2013 for a recent compilation in the Galaxy).  In highly-turbulent
merger galaxies, however, the mixture percentages are likely to be
rather different, especially if the observing beam takes an average
over kpc scales, as it does for galaxies at high redshift.  For a
highly turbulent filament in Stephan's Quintet, Appleton et al.\
(2013) estimate that the standard PDR contribution is at most 10 to
15\% of the C$^+$ line emission.  The large majority of the C$^+$ line
emission is coming from collisional excitation of the ionized fraction
of carbon within the bulk of the warm
molecular gas.

In our picture, on the kpc-scale of our beam, the C$^+$, CO, and dust
emission all come from the same volume.  Carbon within the molecular
clouds is partially ionized by the cosmic rays in the large starburst.
The C$^+$ line itself is excited by collisions with hydrogen molecules,
and its excitation temperature becomes the same as
the kinetic temperature of these collision
partners.  This picture explains why, at least on kpc scales,
the C$^+$ profiles look about the same as the CO profiles:
their linewidths, their velocity gradients, and their excitation, by
the H$_2$ molecules, are the same.

In HDF\,850.1, as explained in the previous sections,
the observed line and continuum
brightness temperatures imply high C$^+$ line optical depth, ``low''
C$^+$ excitation temperature, of the same order as the dust
temperature, with the dust approaching optical thickness at
158\,$\mu$m.  Our interpretation is thus different from the
optically-thin C$^+$ model of Mashian et al. (2013), in which the
C$^+$ and CO lines are emitted from separate regions of the molecular
clouds, with the C$^+$ in a ``high-temperature'' (500\,K) region.  Our
model is in better agreement with their preferred, uniformly mixed CO
and C$^+$ region, although we prefer a somewhat lower kinetic temperature,
which is in better agreement with the dust temperatures inferred from
standard ULIRG and SMG templates.

\section{The $M$(H$_2$)/$L^\prime$(C$^+$) conversion factor}
Ever since the pioneering days of far-IR airborne astronomy, it has
been known that there is a good correlation between the integrated
intensities of the C$^+$ and CO(1--0) lines, and therefore that C$^+$
is a tracer of the molecular gas, and not of atomic hydrogen (e.g.,
Crawford et al.\ 1985; Wolfire et al.\ 1989; Stacey et al.\ 1991).
This relation was observed to hold for both Galactic and extragalactic
sources, including starburst galaxies.  In these earlier papers, the
integrated C$^+$ intensities were given in flux units
(erg\,s$^{-1}$\,cm$^{-2}$\,sr$^{-1}$), with a typical
C$^+$-to-CO(1--0) luminosity ratio ranging from 4400 to 6300.  For 7
galaxies at $z$ = 1 to 2 measured in both C$^+$ and CO, Stacey et
al.\ (2010) found a median ratio of 4400, with a variation of about a
factor of two, up and down.  In their samples (Stacey et al. 1991a; 2010),
cooler, less active, non-starburst galaxies
had much lower C$^+$-to-CO(1--0) ratios.
From another sample of 10 galaxies from
$z$ = 2 to 4, Swinbank et al.\ (2012) obtain a ratio of 4400$\pm
1000$.

In flux
units, the integrated C$^+$ intensities (or line luminosities) contain a
factor of frequency $\nu$, from the conversion from \kms\ to Hz, and
another factor of $\nu^2$, from the conversion from Rayleigh-Jeans
brightness temperature to flux density.
If however, we quote line luminosities in brightness temperature units
(\Kkmspc ), then we must divide these earlier C$^+$-to-CO(1--0)
luminosity ratios by the cube of the frequency ratio, or
(1900.5369\,GHz /115.2712\,GHz)$^3$, which is 4482.
From the observed ratios quoted above, this means that
the C$^+$-to-CO(1--0) luminosity ratio, when expressed in brightness
temperature units, is close to unity, i.e., to within a factor about two,
the $L^\prime$(C$^+$\,158\,$\mu$m) $\approx L^\prime$(CO(1--0)).

Because of this approximate equality of the C$^+$ and  CO(1--0)
luminosities, and because the CO(1--0) luminosity is roughly proportional to the
molecular gas mass, we can therefore use the C$^+$ luminosity to estimate
the molecular gas mass, with nearly the same conversion factor as for
CO(1--0).  The assumptions and justification for using a C$^+$ conversion
factor are thus the same as for using a CO conversion factor.

This is fortunate for us, because even though we do not yet have
sub-arcsecond maps of HDF\,850.1 in any lines of CO, we can
nevertheless estimate the molecular gas mass in the North and South
galaxies directly from their C$^+$ luminosities.  For this, we adopted
a ULIRG conversion factor of 0.8\,\Msun\,(\Kkmspc)$^{-1}$ (Downes \&
Solomon 1998). That factor was based on kinematic models for ULIRGs,
that used escape-probability radiative transfer models for each 3D pixel
of a turbulent, circumnuclear region, in the gravitational
potential of a nuclear stellar bulge.
The calibration was fourfold:  1) matching the calculated
maps of CO intensities to well-resolved observed maps of CO intensities,
2) making the velocity dispersions due to turbulence and due to ordered
motion match the observations in the correct proportions,
3) ensuring that the integrated gas
mass was less than the dynamical mass, and 4) checking that the dust
mass, estimated from the gas mass with a standard gas-to-dust ratio,
was equal to the dust mass estimated directly from observations of the dust
continuum.  No assumptions were made about individual clouds' self-gravity
or cloud counting.   A similar conversion factor has been derived
by Mashian et al.\ (2013), from
LVG fits to the HDF\,850.1 CO and C$^+$ lines observed by Walter et al.\ (2012).

After applying this conversion factor to HDF\,850.1,
the resulting gas masses, corrected for lens magnification
(Table~2), are 1.5 and $0.8\times 10^{10}$\,\Msun\ for the North and
South galaxies respectively.  Our mass of $2.3\times 10^{10}$\,\Msun\
in total, as determined from the C$^+$ luminosity, is therefore in
excellent agreement with the global gas mass estimate (uncorrected
for lensing)
of
$3.5\times 10^{10}$\,\Msun\ derived by Walter et al.\ (2012) by
extrapolating to CO(1--0) from measured higher-$J$ CO lines.
The fact that these two estimates agree so well indicates that there
is not a large contribution to the C$^+$ line from ``CO-dark'' molecular
gas.

\begin{figure} \centering
\includegraphics[angle=-0,width=8.5cm]{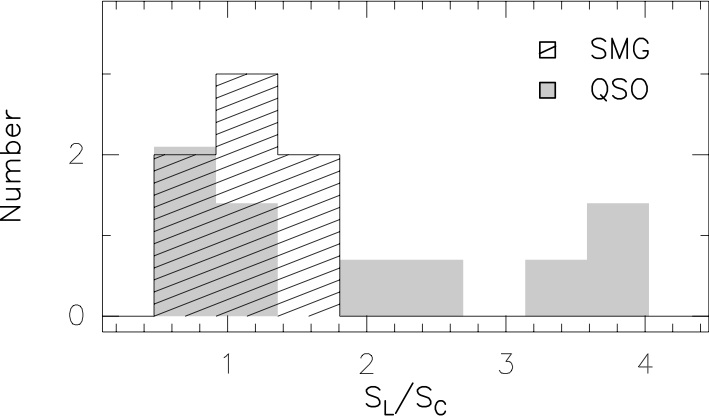}
\caption[Number of sources vs. line-to-continuum ratio]{
Number of sources vs. C$^+$ line-to-continuum ratio, for the sources
listed in Table~3.  The number of quasar-host sources (QSOs) has been normalized
to the number of submillimeter galaxies (SMGs)
so that the areas of the two histograms are the same.
}
\label{Number of sources vs. line-to-continuum ratio}
\end{figure}

\section{Why there is no optical or UV counterpart to HDF\,850.1}
As in many other high-$z$ ULIRGs and SMGs powered by starbursts,
we see directly from the CO, C$^+$, and submillimeter dust luminosities
that HDF\,850.1 contains large quantities of carbon, oxygen, and dust, presumably
ejected by the supernovae in starbursts.
The large mass (typically $10^8$\,\Msun) of
dust may explain why there is no
optical counterpart of HDF\,850.1 in the Hubble Deep Field optical
images.

Furthermore, the relatively
high brightness temperature of the dust continuum from HDF\,850.1 shows
that this large quantity of dust is
nearly opaque at a rest wavelength of 158\,$\mu$m (see section~4).
So if the dust is already opaque in the
far-IR, it will certainly give a complete blackout of any Lyman $\alpha$
that is emitted inside the dusty merger companion galaxies.

Although the submm-emitting dust has such a high optical depth that it
will completely block the optical and UV, some Ly$\alpha$ emission may
not be covered by this dust, and will be able to escape, or,
alternatively, some Ly$\alpha$ may come from another object outside
the submm dust source.  An example of this proximity effect, at
$z=4.7$, is the SMG BR1202$-$07-NW.  No Ly$\alpha$ (or rest-frame
near-IR, or visible light) is detected from this dusty SMG, but
Ly$\alpha$ is definitely detected quite nearby (within $1''$), from
non-SMG companions in the merging group (see, e.g., Salom\'e et al.\
2012, Carilli et al.\ 2013, Carniani et al.\ 2013).

We can also extrapolate from the Chapman et al.\ (2005) detections of
Ly$\alpha$ emission in SMGs.  In their SMG sample, the
Calzetti-extinction-corrected UV luminosities underestimate the true
bolometric luminosities (from the FIR) by a median factor of 120,
which implies that less than one per cent of the UV can
escape. Chapman et al.\ (2005) also give the median redshift of the
SMGs with detected Ly$\alpha$ emission lines as $z = 2.3$.  The
Ly$\alpha$ flux densities in their Fig.2 spectra are typically 2 to
20\,$\mu$Jy. Because flux density decreases as $(1+z)^3$, if these
SMGs with median $z$ = 2.3 already existed at $z=5.2$ with exactly the
same luminosities and emitting areas as at $z$ = 2.3, then their flux
densities would be lower by a factor of $(6.2/3.3)^3 \approx 7$. This
would make many of them marginally detectable at $z=5.2$, at the noise
level in the Ly$\alpha$ spectra of Chapman et al.\ (2005).  If in
addition, these SMGs were to follow the same size decrease as non-SMG
galaxies, with a (1+$z$) scaling (e.g., Oesch et al. 2010; Jiang et
al. 2013), then their Ly$\alpha$ flux densities might be scaled down
by as much as $(1+z)^5$, which would make most of the Chapman et al.\
$z$ = 2.3 sample undetectable at $z=5.2$.   As yet however,
there is no evidence that SMGs follow the same trend of decreasing
size at higher redshift as for non-SMG galaxies.

These three effects, the blockage by the heavy dust by a median
factor of 120 as observed by Chapman et al.\ (2005), the additional
downward scaling in flux density by the usual factor of $(1+z)^3$ due
to the expansion of the universe, and a possible further downward
scaling by $(1+z)^2$, due to smaller intrinsic sizes at higher $z$,
could all contribute to making an optical or UV counterpart to
HDF\,850.1 difficult to detect.

\section{Conclusion: HDF\,850.1 is similar to other
high-$z$ ULIRG and SMG mergers}
To conclude, we summarize how the data in this paper provide evidence that HDF\,850.1 is a merger of two galaxies:

\noindent
1) In the integrated line emission (Fig.~3, left), the source has an
irregular V-like shape, resembling those of interacting systems at
both low and high redshift, rather than a single, undisturbed disk
galaxy.

\noindent
2) The red- and blue-shifted emission (Fig.~1) is concentrated in two
distinct spatial locations, suggesting two merging galaxies.
rather than parts of a large rotating disk.
Instead, the two objects in the C$^+$ maps, resemble
the kpc-sized hotspots observed in the rest-frame optical in $z$ = 2
SMGs by Men\'endez-Delmestre et al.\ (2013), which those authors also interpret
as merger remnants, due to the lack of any spectroscopic
kinematic signatures of rotating disks.

\noindent
3) The line profiles of the red- and blue-shifted components (Fig.~2,
upper and middle) are roughly Gaussian, with large linewidths
($\sim$300 and 410\,\kms ), characteristic of individual galaxies,
rather than the linewidths expected from a
large, quiescent disk galaxy.

\noindent
4) The total velocity spread of 940\,\kms\ in HDF\,850.1 is typical of
ULIRG mergers, like Arp~220, IRAS\,17208-0014, Mrk273, NGC6240, and others.
It is difficult to explain such a large velocity spread with a
rotating disk, at any redshift.

\noindent
5) HDF\,850.1 has a high IR luminosity of $6\times 10^{12}$\,\Lsun\
(Table~2, and Walter et al.\ 2012).  This is comparable with the IR
luminosities of those low-$z$ ultraluminous galaxies (ULIRG) that are
known to be mergers.  It is also comparable with high-$z$
submillimeter galaxies, for which most of the bright SMGs with
$L_{IR}>5\times 10^{12}$\,\Lsun\ are also known to be major mergers
(e.g., Engel et al.\ 2010).  Further evidence that many of the
high-$z$ IR luminous galaxies are mergers rather than disks is the
recent study by Hung et al.\ (2013) of 2084 {\it Herschel}
far-IR-selected galaxies at $0.2<z<1.5$ in the COSMOS field.  Their
study shows that the fraction of disk galaxies decreases with
increasing $L_{IR}$, and that in the highest IR luminosity bins
($>10^{11.5}$\,\Lsun ), more than half of the objects have
interacting/merger morphologies, rather than disk galaxy morphologies.

\noindent
6) It is not only the high FIR luminosities that argue for a merger,
but also the high CO line luminosities of $4\times 10^{10}$\,\Kkmspc\
(Walter et al.\ 2012), that are characteristic of ULIRG merger
galaxies, and are at least an order of magnitude greater than those of
disk galaxies in the local universe.
The fact that, in brightness temperature units, $L^\prime$(C$^+$) $\approx L^\prime$(CO(1--0)),
shows that HDF\,850.1 is definitely in the starburst class,
rather than in the non-starburst class of galaxies that have much lower
C$^+$-to-CO ratios (Stacey et al.\ 1991; 2010). We interpret the ULIRG-level
HDF\,850.1 starburst as
being induced by the merger.

\noindent
7) The absence of an optical counterpart suggests high dust
obscuration, typical of the gas-rich mergers like Arp~220 and other
ULIRG mergers.  (see discussion in the previous section).  The optical
depth of the dust radiation is close to unity at rest-frame
158\,$\mu$m.  This is highly unusual.  It is not observed for any
known disk galaxies, but it is observed in the highly concentrated
dust in merging galaxies, such as Arp~220 (e.g., Fischer et al. 1997;
Downes \& Eckart 2007).

\noindent
8) HDF\,850.1 is the only known SMG in the galaxy overdensity at $z$ = 5.2
   (Walter et al. 2012), suggesting a rare,
   transient event within this galaxy group, namely a
   very large starburst, characteristic of gas-rich mergers.

\noindent
9) The linear sizes of the two objects in HDF\,850.1 (Table~2) are
about the same as the 1 to 2\,kpc sizes that are measured for
individual galaxies themselves at $z$ = 5 (e.g., Oesch et al. 2010;
Jiang et al. 2013).  This also suggests that rather than parts of a
large disk, they are pieces in the merger assembly of what will be a
single, larger galaxy at later times.

\noindent
10) The classic argument against the existence of large disks at $z$ =
5 is the angular momentum argument (see, e.g., Rees 1995).  Analytical
and numerical estimates indicate that when protogalaxies decouple from
the expansion of the universe, tidal torquing between them would
impart $<10$\% of the rotation needed for centrifugal support. Hence
for a large ($\sim$10\,\,kpc ) rotationally-supported disk to form,
the infalling disk material must have acquired its angular momentum at
distances $>10^5$\,pc.  The infall timescale from such distances
however, is comparable with the age of the universe at $z$ = 5.2.
This means that large, settled-down, 10\,kpc-scale galaxy disks cannot
be in place much before $z$ = 2. Our size measurements of the
HDF\,850.1 North and South objects are consistent with these ideas, as
are the recent measurements of kpc-sized rotating molecular disks in
BR1202-07-NW and SE at $z$ = 4.7 (Salom\'e et al.\ 2012, Carilli et
al.\ 2013, Carniani et al.\ 2013).  The infalling gas does have time
to establish rotational support in kpc-scale disks by $z$ = 5, but not
in the order-of-magnitude larger disks that may be in place by $z$ =
2.  By this angular momentum argument, it is rather unlikely that the
entire structure, consisting of the HDF\,850.1 North and South objects
together, is one large, rotating disk at $z$ = 5.2.

\begin{acknowledgements}
We thank the Plateau de Bure Interferometer operators for their help
with the observing, and the referee for helpful comments and suggestions for
improving the text.
IRAM is supported by INSU/CNRS (France), MPG
(Germany) and IGN (Spain).
\end{acknowledgements}


\newpage

\end{document}